\documentclass[doublecol]{epl2}

\usepackage{amsfonts}
\usepackage{amssymb}
\title{Self-organization of price fluctuation distribution in evolving
markets}
\shorttitle{Price fluctuations in evolving markets}
\author{Raj Kumar Pan \and Sitabhra Sinha\footnote{E-mail:
\email{sitabhra@imsc.res.in}} }
\institute{%
The Institute of Mathematical Sciences, C.I.T. Campus, Taramani,
Chennai - 600 113 India
}%
\date{\today}


\abstract{
Financial markets can be seen as complex systems in non-equilibrium steady
state, one of whose most important properties is the distribution of price
fluctuations. Recently, there have been assertions that this distribution
is qualitatively different in emerging markets as compared to developed
markets.  Here we analyse both high-frequency tick-by-tick as well as daily
closing price data to show that the price fluctuations in the Indian stock
market, one of the largest emerging markets, have a distribution that is
identical to that observed for developed markets (e.g., NYSE).  In
particular, the cumulative distribution has a long tail described by a
power law with an exponent $\alpha \approx 3$.  Also, we  study the
historical evolution of this distribution over the period of existence of
the National Stock Exchange (NSE) of India, which coincided with the rapid
transformation of the Indian economy due to liberalization, and show that
this power law tail has been present almost throughout.  We conclude that
the ``inverse cubic law'' is a truly universal feature of a financial
market, independent of its stage of development or the condition of the
underlying economy.}

\begin{document}
\maketitle

\section{Introduction}
Financial markets are paradigmatic examples of complex systems, comprising
a large number of interacting components that are subject to a constant
flow of external information~\cite{Stanley99,Bouchaud03}. Statistical
physicists have studied simple interacting systems which self-organize into
non-equilibrium steady states, often characterized by power law
scaling~\cite{Privman97}.  Whether markets also show such behavior can be
examined by looking for evidence of scaling functions which are invariant
for different markets.  The most prominent candidate for such an universal,
scale-invariant property is the cumulative distribution of stock price
fluctuations. The tails of this distribution has been reported to follow a
power law, $P_{c}(x) \sim x^{-\alpha}$, with the exponent 
$\alpha \approx 3$~\cite{Gopikrishnan98}.  
This ``inverse cubic law'' had been reported
initially for a small number of stocks from the S\&P 100
list~\cite{Jansen91}.  Later, it was established from statistical analysis
of stock returns in the German stock exchange~\cite{Lux96}, as well as for
three major US markets, including the New York Stock Exchange
(NYSE)~\cite{Plerou99}.  The distribution was shown to be quite robust,
retaining the same functional form for time scales of upto several days.
Similar behavior has also been seen in the London Stock
Exchange~\cite{Farmer04}.  An identical power law tail has also been
observed for the fluctuation distribution of a number of market
indices~\cite{Gopikrishnan99,Oh06}.  This apparent universality of the
distribution may indicate that different markets self-organize to an almost
identical non-equilibrium steady state.  However, as almost all these
observations are from developed markets, a question of obvious interest is
whether the same distribution holds for developing or emerging financial
markets. If the inverse cubic law is a true indicator of self-organization
in markets, then observing the price fluctuation distribution as the market
evolves will inform us about the process by which this complex system
converges to the non-equilibrium steady state characterizing developed
markets.

However, when it comes to empirical reports about such emerging markets
there seems to be a lack of consensus. The market index fluctuations in
Brazil~\cite{Couto01} and Korea~\cite{Oh06} have been reported to follow an
exponential distribution, while, the distribution for an Indian market
index was observed to be heavy tailed with exponent greater then
3~\cite{Sarma05}.  On the other hand, a comparative analysis of 27 indices
from both mature and emerging markets found their tail behavior to be
similar~\cite{Jondeau99}.  It is hard to conclude about the nature of the
fluctuation distribution for individual stock prices from the index data,
as the latter is a weighted average of several stocks. Therefore, in
principle, the index can show a distribution quite different from that of
its constituent stocks if their price movements are not correlated.
\begin{figure*}
\begin{center}
\includegraphics[width=0.87\linewidth]{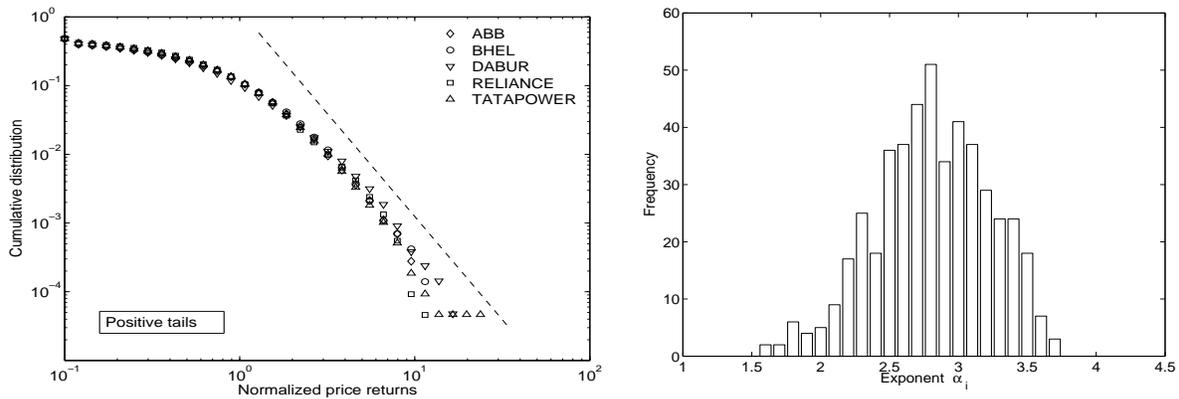}
\caption{(Left) Cumulative distribution of the positive tails of the
normalized 5-min returns distribution of 5 stocks chosen arbitrarily from
those listed in the NSE for the period January 2003 to March 2004. The broken
line indicates a power law with exponent $\alpha = 3$.  (Right) The
histogram of the power-law exponents obtained by regression fit for the
positive tail of individual cumulative return distributions of 489 stocks.
The median of the exponent values is 2.84.} 
\label{comp_5min}
\end{center}
\end{figure*}

Analysis of individual stock price returns for emerging markets have also
not resulted in an unambiguous conclusion about whether such markets behave
differently from developed markets.  A study of the fluctuations in the
daily price of the 49 largest stocks in an Indian stock exchange has
claimed that the distribution has exponentially decaying
tails~\cite{Matia04}.  This implies the presence of a characteristic scale,
and the breakdown of universality of the power law tail for the price
fluctuation distribution. On the other hand, it has been claimed that this
distribution in emerging markets has even more extreme tails than developed
markets, with an exponent $\alpha$ that can be less than
$2$~\cite{Bouchaud05}.  Recently, there has been a report of the ``inverse
cubic law'' for the daily return distribution in the Chinese stock markets
of Shanghai and Shenzhen~\cite{Gu06}. These contradictory reports indicate
that a careful analysis of the stock price return distribution for emerging
markets is extremely necessary. This will help us to establish definitively
whether the ``inverse cubic law'' is invariant with respect to the stage of
economic development of a market.

All the previous studies of price fluctuations in emerging markets have
been done on low-frequency daily data. For the first time, we report
analysis done on high-frequency tick-by-tick data, which are corroborated
by analysis of daily data over much longer periods. The data set that we
have chosen for this purpose is from the National Stock Exchange (NSE) of
India, the largest among the 23 exchanges in India, with more than 85\% of
the total value of transactions for securities in all market segments of
the entire Indian financial market in recent times \cite{Nse}.  This data
set is of unique importance, as we have access to daily data right
from the time
the market commenced operations in the equities market in November 1994,
upto the present when it has become the world's third largest stock
exchange (after NASDAQ and NYSE) in terms of transactions \cite{FIBV}.
Over this period, the market has grown rapidly, with the number of
transactions having increased by more than three orders of magnitude.
Therefore, if markets do show discernible transition in the return
distribution during their evolution, the Indian market data is best
placed to spot evidence for it, not least because of the rapid transformation
of the Indian economy in the liberalized environment since the 1990s.

In this paper, we focus on two important questions: (i)~Does
an emerging market exhibit a different price fluctuation distribution
compared to developed markets, and (ii)~if the market is indeed following
the inverse cubic law at present, whether this has been converged at
starting from an initially different distribution when the market had
just begun operation. Both of these questions are answered in the
negative in the following analysis. 

\section{Data description}
We have looked at two data sets having different temporal resolutions: (i)
The high-frequency tick-by-tick data contains information about all
transactions carried out in the NSE between January 2003 and March 2004.  This
information includes the date and time of trade, the price of the stock
during transaction and the volume of shares traded. This database is
available in the form of CDs published by NSE. For calculating the price
return, we have focused on 489 stocks that were part of the BSE 500 index
(a comprehensive indicator for the Indian financial market)
during this period.  The number of transactions for each company in this
set is $\sim 10^{6}$, on the average.  The total number of transactions for
the 489 stocks is of the order of $5 \times 10^8$ during the period under
study.  (ii) The daily closing price of all the stocks listed in
NSE during its period of existence between November 1994 and May 2006. This was
obtained from the NSE website~\cite{Nse_web}
and manually corrected for stock splitting.  For comparison with
US markets, in particular the NYSE, we have considered the 500 stocks
listed in S\&P 500 during the period November 1994 - May 2006, the daily data
being obtained from Yahoo! Finance \cite{Yahoo}.
\begin{figure*}
\begin{center}
  \includegraphics[width=0.9\linewidth]{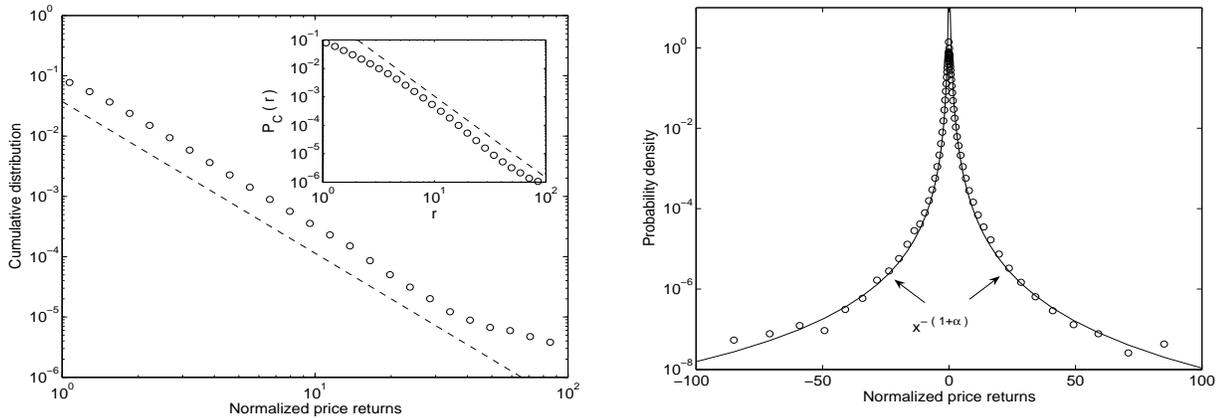}
\caption{(left) Cumulative distribution of the negative and (inset)
positive tails of the normalized returns for the aggregated data of 489 
stocks in the NSE for the period January 2003 to March 2004. The
broken line for visual guidance indicates the power law asymptotic form.
(Right) Probability density function of the normalized returns. The solid
curve is a power-law fit in the region $1-50$. We find that the
corresponding cumulative distribution exponent, $\alpha=2.87$ for the
positive tail and $\alpha=2.52$ for the negative tail.}
\label{tail_5min}
\end{center}
\end{figure*}

\section{Results}
To measure the price fluctuations such that the result is independent of
the scale of measurement, we calculate the logarithmic return of price.  If
$P_{i}(t)$ is the stock price of the $i$th stock at time $t$, 
then the (logarithmic) price return is defined as 
\begin{equation}
R_{i}(t,\Delta t) \equiv \ln {P_{i}(t+\Delta t)}- \ln {P_{i}(t)}.
\label{return}
\end{equation}
However, the distribution of price returns of different stocks may
have different widths, owing to differences in their volatility, defined
(for the $i$-th stock) as $\sigma_{i}^{2} \equiv \langle R_{i}^{2} \rangle
- \langle R_{i} \rangle^{2}$.  To compare the distribution of different
stocks, we normalize the returns by dividing them with their volatility
$\sigma_{i}(t)$ as in Ref~\cite{Bouchaud03}.
The resulting normalized price return 
\footnote{The normalization of return $R_i (t)$ is performed by 
removing its own contribution from the volatility, i.e., 
$\sigma_i (t)=\sqrt{\frac{1}{N-1} \sum_{t^\prime \neq t} \{R_i
(t^{\prime})\}^2 
-\langle R_i (t) \rangle}^2$.}
is given by
\begin{equation} 
r_{i}(t,\Delta t) \equiv \frac{R_{i} (t) -\langle R_{i} (t) \rangle}{\sigma_{i} (t) },
\label{normalized_return}
\end{equation}
where $\langle \cdots \rangle$ denotes the time average over the given
period. 

For analysis of the high-frequency data, we consider the aforementioned
489 stocks.
Choosing an appropriate $\Delta t$, we obtain the corresponding return by
taking the log ratio of consecutive average prices, averaged over a time
window of length $\Delta t$. Fig.~\ref{comp_5min}~(left) shows the
cumulative distribution of the normalized returns $r_{i}$ with $\Delta t =
5$ mins for five stocks, arbitrarily chosen from the dataset. We observe
that the distribution of normalized returns $r_{i}$ for all the stocks have
the same functional form with a long tail that follows a power-law
asymptotic behavior. The distribution of the corresponding power law
exponent $\alpha_{i}$ for all the 489 stocks that we have considered is
shown in Fig~\ref{comp_5min}~(right).

As all the individual stocks follow very similar distributions, we can
merge the data for different stocks to obtain a single distribution for
normalized returns.  The aggregated return data set with $\Delta t =5$ mins
has $6.5 \times 10^6$ data points. The corresponding cumulative
distribution is shown in Fig.~\ref{tail_5min}~(left), with the exponents
for the positive and negative tails estimated as 
\begin{equation}
\alpha = \left\{ \begin{array}{ll}
2.87 \pm 0.08 & \mbox{(positive tail)} \\
2.52 \pm 0.04 & \mbox{(negative tail).}
\end{array} \right.
\label{eqn:tail_exponent}
\end{equation}
From this figure we confirm that the distribution does indeed follow a
power law decay, albeit with different exponents for the positive and
negative return tails. Such a difference between the positive and negative
tails have also been observed in the case of stocks in the
NYSE~\cite{Plerou99}. To further verify that the tails are indeed
consistent with a power law form, we perform an alternative measurement of
$\alpha$ using the Hill estimator~\cite{Hill75,Dress00}.  We arrange the
returns in decreasing order such that $r_{1}>\cdots>r_{n}$ and 
obtain the Hill estimator (based on the largest $k+1$ values) as
$ H_{k,n}=\frac{1}{k}\sum_{i=1}^{k}{\log \frac{r_{i}}{r_{k+1}}},$
for $k=1,\cdots,n-1$. The estimator $H_{k,n} \rightarrow \alpha^{-1}$ when
$k \rightarrow \infty$ and $k/n \rightarrow 0$.  For our data, this
procedure gives $\alpha = 2.86$ and $2.56$ for the positive and the
negative tail respectively (when $k=20,000$), which are consistent 
with~(\ref{eqn:tail_exponent}).  
\begin{figure*}
\begin{center}
  \includegraphics[width=0.87\linewidth]{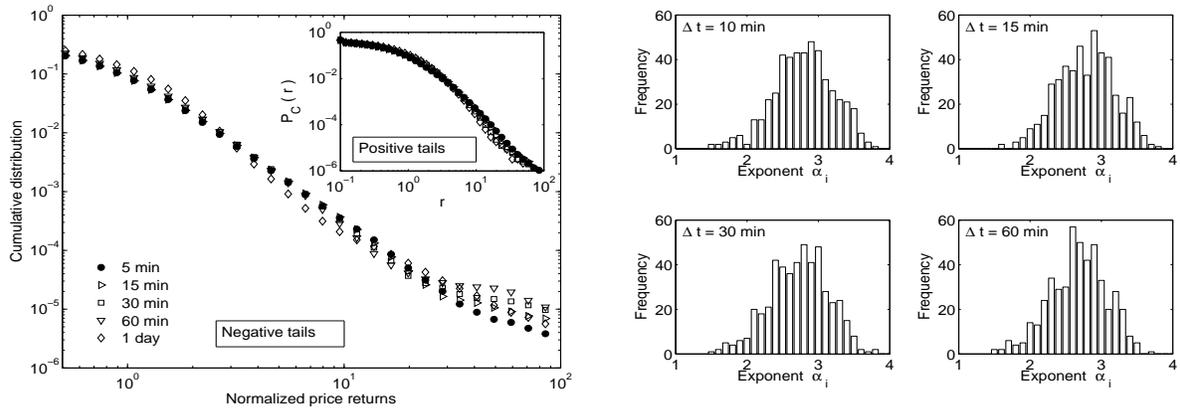}
\caption{(Left) Cumulative distribution of the negative and (inset) positive 
tails of the normalized returns distribution for different time scales 
($\Delta t \leq 1$ day). (Right) Histograms 
of the power-law exponents for each of the 489 stocks, obtained by regression 
fit on the
positive tail of cumulative return distributions,
for different time scales (10 min $\leq \Delta t \leq 60$ min).}
\label{diff_time}
\end{center}
\end{figure*}

Next, we extend this analysis for longer time scales, to observe how the
nature of the distribution changes with increasing $\Delta t$. As has been
previously reported for US markets, the distribution is found to decay
faster as $\Delta t$ becomes large. However, upto $\Delta t$ = 1 day, i.e.,
the daily closing returns, the distribution clearly shows a power-law tail
(Fig.~\ref{diff_time},~left). The deviation is because of the decreasing
size of the data set with increase in $\Delta t$.
Note that, while for $\Delta t<1$ day we have used
the high-frequency data, for $\Delta t=1$ day we have considered the longer
data set of closing price returns for all stocks traded in NSE between
November 1994 to May 2006. 
In Fig.~\ref{diff_time}~(right) we have also shown the distributions of the
power-law exponents for the individual stocks, for 
10 min $\leq \Delta t \leq 60$ min. We observe that the bulk
of the exponent falls between 2 and 4, consistent with the results from
the merged data sets.
\begin{figure}
\begin{center}
  \includegraphics[width=0.85\linewidth]{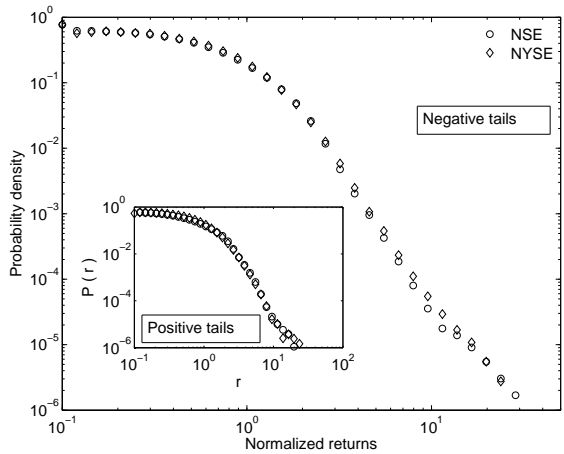}
\caption{Comparison of the negative and (inset) positive tails of the 
normalized daily returns distribution 
for all stocks traded at NSE ($\circ$) and 500 stocks traded at NYSE
($\diamond$) during the period November 1994 to May 2006.}
\label{nyse_nse}
\end{center}
\end{figure}

To compare the distribution of returns in this emerging market with that
observed in mature markets, we have considered the daily return data
for the 500 stocks from NYSE listed in S\&P 500 over the same period. As
seen in Fig.~\ref{nyse_nse}, the distributions for NSE and NYSE are almost
identical, implying that the price fluctuation distribution of emerging
markets cannot be distinguished from that of developed markets, contrary to
what has been claimed recently~\cite{Matia04}.
\begin{figure}
\begin{center}
  \includegraphics[width=0.85\linewidth]{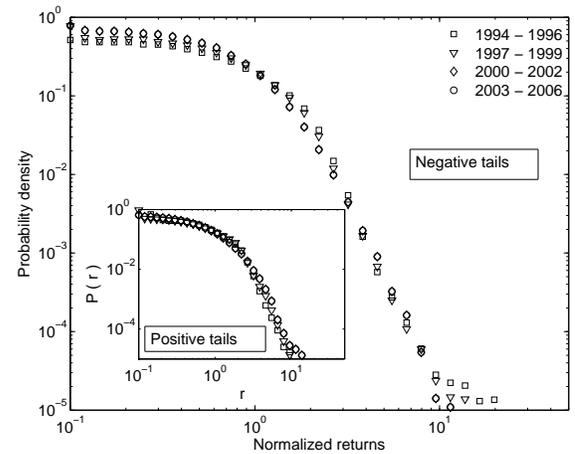}
\caption{The negative and (inset) positive tails of the 
normalized daily returns distribution for
all NSE stocks traded during the periods
1994-1996 ($\square$), 1997-1999 ($\triangledown$), 2000-2002 ($\diamond$) 
and 2003-2006 ($\circ$).}
\label{diff_period}
\end{center}
\end{figure}

We now turn to the second question, and check whether it is possible to see
any discernible change in the price fluctuation
distribution as the stock market evolved over time. 
For this we focus on the daily return distribution for
{\em all} stocks that were traded during 
the entire period of existence of NSE. This period is
divided into four intervals (a) 1994-1996,
(b) 1997-1999, (c) 2000-2002, and (d) 2003-2006~\footnote{Total number of 
stocks traded in these four intervals were 
1460, 1560, 1321 and 1160 respectively.
}, each
corresponding to increase in the number of transactions by
an order of magnitude.
Fig.~\ref{diff_period} shows that the return distribution at all four
periods are similar, the negative tail even more so than the positive one.
While the numerical value of the tail exponent may appear to have changed
somewhat over the period that the NSE has operated,
the power law nature of the tail is apparent at even the
earliest period of its existence.  We therefore conclude that the
convergence of the return distribution to a power law functional form is
extremely rapid, indicating that a market is effectively always at the
non-equilibrium steady state characterized by the inverse cubic law.

We have also verified that stocks in the Bombay Stock Exchange (BSE),
the second largest in India after NSE, follow
a similar distribution~\cite{Sinha06}. Moreover, the return distribution of
several Indian market indices (e.g., the NSE Nifty) also exhibit power law
decay, with exponents very close to 3~\cite{Pan06}.  As the index is a
composite of several stocks, this behavior can be understood as a
consequence of the power law decay for the tails of individual stock price
returns, provided the movement of these stocks are
correlated~\cite{Plerou99,Sinha06}. 
Even though the Indian market microstructure 
has
been refined and modernized significantly in the period under study as a
result of the reforms and initiatives taken by the government, 
the nature of
the return distribution has remained invariant, indicating that the nature
of price fluctuations in financial markets is most probably independent of
the level of economic development.

\section{Discussion and Conclusion}
Most of the previous studies on emerging markets had focussed on either
stock indices or a small number of stocks. In addition, all these studies
were done with low-frequency daily data. Thus, the number of data points
used for calculating the return distribution were orders of magnitude
smaller compared to ours. Indeed, the paucity of data can result in missing
the long tail of a power law distribution and falsely identifying it to be
an exponential distribution.  Matia et al~\cite{Matia04} claimed that
differences in the daily return distribution for Indian and US markets were
apparent even if one looks at only 49 stocks from each market. However, we
found that this statement is critically dependent upon the choice of
stocks.  Indeed, when we made an arbitrary choice of 50 stocks in both
Indian and US markets, and compared their distributions, we found them to
be indistinguishable. Therefore, the results of analysis done on such small
data sets can hardly be considered stable, with the conclusions depending
on the particular sample of stocks.

In this study, we have shown conclusively that the inverse cubic law for
price fluctuations holds even in emerging markets.  It is indeed surprising
that the nature of price fluctuations is invariant with respect to large
changes in the number of stocks, trading volume and number of transactions
that have all increased significantly at NSE during the period under study.
The robustness of the distribution implies that it should be possible to
explain it independent of the particular features of different markets, or
the various economic factors underlying them.

\acknowledgements
We are grateful to M.~Krishna for assistance in obtaining and
analyzing the high-frequency NSE data. We thank A.~Chatterjee, J.D.~Farmer and
the referees for helpful comments.

\end{document}